\RequirePackage{fix-cm}

\documentclass[a4paper,reqno,12pt]{amsart}
\usepackage[text={5.9in,8.5in},centering,top=30mm]{geometry}

\DeclareSymbolFont{cmsymbols}{OMS}{cmsy}{m}{n}
\SetSymbolFont{cmsymbols}{bold}{OMS}{cmsy}{b}{n}
\DeclareSymbolFontAlphabet{\mtc}{cmsymbols}

\DeclareSymbolFont{mypnc}{OML}{pnc}{m}{n}
\DeclareMathSymbol{\pazotau}{\mathalpha}{mypnc}{"1C}
\DeclareMathSymbol{\pazoomn}{\mathalpha}{mypnc}{"21}

\DeclareMathSymbol{\pazoCZ}{\mathalpha}{mypnc}{"5A}

\usepackage{amsbsy}

\makeatletter
\newsavebox\myboxA
\newsavebox\myboxB
\newlength\mylenA

\newcommand*\xoverline[2][0.75]{%
    \sbox{\myboxA}{$\m@th#2$}%
    \setbox\myboxB\null
    \ht\myboxB=\ht\myboxA%
    \dp\myboxB=\dp\myboxA%
    \wd\myboxB=#1\wd\myboxA
    \sbox\myboxB{$\m@th\overline{\copy\myboxB}$}
    \setlength\mylenA{\the\wd\myboxA}
    \addtolength\mylenA{-\the\wd\myboxB}%
    \ifdim\wd\myboxB<\wd\myboxA%
       \rlap{\hskip 0.5\mylenA\usebox\myboxB}{\usebox\myboxA}%
    \else
        \hskip -0.5\mylenA\rlap{\usebox\myboxA}{\hskip 0.5\mylenA\usebox\myboxB}%
    \fi}
\makeatother

\usepackage{graphicx}
\usepackage{mathptmx}

\usepackage{times}

\usepackage{bm}

\usepackage{amsmath,amsthm,amssymb}

\usepackage{enumerate}
\usepackage{tabularx}
\usepackage[mathscr]{euscript}
\usepackage{eufrak}

\usepackage{stmaryrd}

\usepackage{pgf}
\usepackage{tikz}
\usetikzlibrary{patterns}

\usepackage[colorlinks=true,
            linkcolor=blue,
            urlcolor=black,
            citecolor=blue]{hyperref}

\providecommand{\D}{\mathbb}

\newcommand{\ii}{\mathrm{i}}

\providecommand{\abs}[1]{\lvert#1\rvert}

\providecommand{\norm}[1]{\lVert#1\rVert}

\providecommand{\prob}[1]{\D{P}\left\{ #1 \right\}}

\newcommand{\minus}{\setminus}

\newcommand{\all}{\forall\,}
\newcommand{\exist}{\exists\,}

\newcommand{\dist}{\mathrm{dist}}
\newcommand{\Ord}[1]{\mathrm{O}\left(#1\right)}
\newcommand{\ord}[1]{\mathrm{o}\left(#1\right)}

\newcommand{\supp}{\mathrm{supp\,}}

\newcommand{\one}{\mathbf{1}}

\def\mcA{\mtc{A}}
\def\mcB{\mtc{B}}
\def\mcC{\mtc{C}}

\def\mcE{\mtc{E}}
\def\mcG{\mtc{G}}
\def\mcH{\mtc{H}}

\def\mcM{\mtc{M}}

\def\mcS{\mtc{S}}

\def\mcX{\mtc{X}}

\def\mcZ{\mtc{Z}}

\def\btau{\xoverline{\pazotau}}
\def\btaukap{\xoverline{\pazotau}_{\!\!\kap}}
\def\ttau{\pazotau}

\newcommand{\fukap}{\overline{\mathfrak{u}}^{(\varkappa)}}

\def\llb{\llbracket}
\def\rrb{\rrbracket}

\newcommand{\obball}{{\overline{\bball}}}

\newcommand{\rb}{\mathrm{b}}

\newcommand{\tW}{\widetilde{W}}

\newcommand{\gea}{\gtrsim}
\newcommand{\lea}{\lesssim}

\newcommand{\hC}{\widehat{C}}

\definecolor{redd}{rgb}{0.95,0.2,0.2}
\definecolor{gris}{rgb}{0.9,0.9,0.9}
\definecolor{greenn}{rgb}{0.1,0.6,0.2}

\definecolor{cmgray}{rgb}{0.7,0.7,0.7}

\definecolor{cmblue}{rgb}{0.2,0.5,0.8}

\newcommand{\be}{\begin{equation}}
\newcommand{\ee}{\end{equation}}
\newcommand{\ba}{\begin{array}{l}}
\newcommand{\ea}{\end{array}}

\newcommand{\bal}{\begin{aligned}}
\newcommand{\eal}{\end{aligned}}

\newcommand{\baln}{\begin{align}}
\newcommand{\ealn}{\end{align}}

\newcommand{\ble}{\begin{lemma}}
\newcommand{\ele}{\end{lemma}}

\newcommand{\bco}{\begin{cor}}
\newcommand{\eco}{\end{cor}}

\newcommand{\bpr}{\begin{proposition}}
\newcommand{\epr}{\end{proposition}}

\newcommand{\bre}{\begin{remark}}
\newcommand{\ere}{\end{remark}}

\newcommand{\btm}{\begin{thm}}
\newcommand{\etm}{\end{thm}}

\newcommand{\bde}{\begin{definition}}
\newcommand{\ede}{\end{definition}}

\newcommand{\lcite}[2]{{\cite[#2]{#1}}}

\newcommand{\myset}[1]{\left\{ \, #1 \, \right\}}

\newcommand{\eu}{{\mathrm{e}}}

\newcommand{\ffi}{\varphi}

\newcommand{\BDelta}{\boldsymbol{\Delta}}

\newcommand{\half}{\frac{1}{2}}

\newcommand{\eps}{\epsilon}

\newcommand{\del}{\delta}

\DeclareSymbolFont{newfont}{OML}{cmm}{m}{it}
\DeclareMathSymbol{\Epsilon}{3}{newfont}{15}
\DeclareMathSymbol{\Varrho}{3}{newfont}{37}
\DeclareMathSymbol{\rro}{3}{newfont}{37}

\newcommand{\pt}{\partial}
\newcommand{\Const}{\mathrm{Const\,}}

\newcommand{\pr}[1]{\mathbb{P}\left\{\vphantom{a^{a}_{p}} #1 \right\}}

\newcommand{\prsub}[2]{\mathbb{P}_{#1}\left\{ #2 \right\}}

\newcommand{\esm}[1]{\D{E}\left[\, #1\, \right]}

\newcommand{\vertii}[1]{{\big\vert\kern-0.25ex\big\vert #1
    \big\vert\kern-0.25ex\big\vert\kern-0.25ex}}

\newcommand{\vertiii}[1]{{\big\vert\kern-0.25ex\big\vert\kern-0.25ex\big\vert #1
    \big\vert\kern-0.25ex\big\vert\kern-0.25ex\big\vert}}

\newcommand{\kap}{\varkappa}
\newcommand{\lam}{\lambda}
\newcommand{\om}{\omega}

\newcommand{\Bfq}{\pmb{\mathfrak{q}}}
\newcommand{\BfQ}{\pmb{\mathfrak{Q}}}

\newcommand{\Lam}{\Lambda}
\newcommand{\Om}{\Omega}

\newcommand{\ket}[1]{|#1\rangle}
\newcommand{\bra}[1]{\langle#1|}
\newcommand{\lr}[1]{\langle#1\rangle}

\newcommand{\bcube}{{\boldsymbol{\Lambda}}}

\newcommand{\ball}{\mathrm{B}}

\newcommand{\fa}{\mathfrak{a}}
\newcommand{\fb}{\mathfrak{b}}

\newcommand{\fc}{\mathfrak{c}}

\newcommand{\fn}{\mathfrak{n}}
\newcommand{\fp}{\mathfrak{p}}
\newcommand{\fr}{\mathfrak{r}}
\newcommand{\fq}{\mathfrak{q}}
\newcommand{\fs}{\mathfrak{s}}

\newcommand{\fB}{\mathfrak{B}}

\newcommand{\fF}{\mathfrak{F}}

\newcommand{\fu}{\mathfrak{u}}

\newcommand{\BF}{\mathbf{F}}

\newcommand{\BG}{\mathbf{G}}

\newcommand{\BH}{\mathbf{H}}

\newcommand{\BU}{\mathbf{U}}
\newcommand{\BV}{\mathbf{V}}

\newcommand{\BPsi}{\boldsymbol{\Psi}}

\newcommand{\cscB}{\mathcal{B}}

\newcommand{\cC}{\mathcal{C}}

\newcommand{\cE}{\mathcal{E}}

\newcommand{\bball}{\mathbf{B}}

\newcommand{\Bx}{\mathbf{x}}

\newcommand{\By}{\mathbf{y}}
\newcommand{\Bz}{\mathbf{z}}
\newcommand{\Bu}{\mathbf{u}}

\newcommand{\pa}[1]{\left( #1 \right)}

\newcommand{\ra}{\mathrm{a}}
\newcommand{\rc}{\mathrm{c}}

\newcommand{\rh}{\mathrm{h}}

\newcommand{\rrq}{\mathrm{q}}

\newcommand{\rP}{\mathrm{P}}

\newcommand{\DC}{\mathbb{C}}
\newcommand{\DP}{\mathbb{P}}
\newcommand{\DR}{\mathbb{R}}

\newcommand{\DZ}{\mathbb{Z}}

\newcommand{\DN}{\mathbb{N}}

\newcommand{\etal}{\emph{et al.}\xspace}

\newcommand{\cond}{\,\big|\,}

\newenvironment{hyp}[1]{
\vskip3mm\par\noindent
$\blacklozenge$\;\textbf{Hypothesis #1.}
\noindent}

\newcommand{\bhy}[1]{\begin{hyp}{#1}}
\newcommand{\ehy}{\end{hyp}}

\newtheorem{thm}{Theorem}[section]
\newtheorem{cor}{Corollary}

\newtheorem{proposition}[thm]{Proposition}

\usepackage{amssymb}
\usepackage{amsmath}
\usepackage{mathrsfs}

\usepackage{xspace}

\usepackage{epsfig}
\usepackage{enumerate}
\usepackage{graphicx}

\usepackage[title]{appendix}

\numberwithin{equation}{section}

\newtheorem{lemma}{Lemma}[section]

\newtheorem{remark}{Remark}[section]

\newtheorem{definition}{Definition}[section]

\def\sigal{$\sigma$-algebra\,}

\newcommand{\set}[1]{\left\{#1\right\}}

\def\tu{\tilde{u}}
\def\tball{\tilde{\ball}}

\def\tBH{\tilde{\BH}}

\def\BW{\mathbf{W}}

\def\bea#1\eea{\begin{align}#1\end{align}}
\def\bea#1\eea{\begin{align}#1\end{align}}

\def\beal{\begin{equation}\begin{aligned}}
\def\eeal{\end{aligned}\end{equation}}

\def\moins{\setminus}

\allowdisplaybreaks

\begin{document}

\title[N-particle Bernoulli--Anderson model]
{Density of States under non-local interactions III. \\N-particle Bernoulli--Anderson model}



\author{Victor Chulaevsky}

\address{Universit\'{e} de Reims\\
D\'{e}partement de math\'{e}matiques\\
51687 Reims Cedex, France}

\email{victor.tchoulaevski@univ-reims.fr}

\date{\today}

\maketitle

\begin{abstract}
Following \cite{C16e,C16f}, we analyze regularity properties of single-site probability distributions of the
random potential and of the Integrated Density of States (IDS) in the Anderson models with infinite-range interactions
and arbitrary nontrivial probability distributions of the site potentials.
In the present work, we study $2$-particle Anderson Hamiltonians on a lattice and prove spectral and strong dynamical
localization at low energies, with exponentially decaying eigenfunctions, for a class of site potentials featuring a
power-law decay.

\end{abstract}


\section{Introduction}
\label{sec:intro}

This text is a follow-up of \cite{C16e}, where the reader can find the main motivations,
a historical review, and a number of bibliographical references.

We consider a $2$-particle lattice Anderson Hamiltonian
\be
\BH(\om) = \BH_0 + \BV(\Bx,\om) + \BU(\Bx),
\ee
where $\BH$ is the kinetic energy operator, which we assume to be the standard second-order lattice Laplacian,
$\BU(\Bx)$ is the inter-particle interaction potential of the form
\be
\BU(\Bx) = \sum_{i<j} U^{(2)}(|x_i - x_j|),
\ee
with a compactly supported two-body interaction potential $U^{(2)}$,
and $\BV(\Bx,\om)$ is the operator of multiplication by the random external potential energy of the form
\be
\BV(\Bx,\om) = \sum_{j=1}^N V(x_j,\om),
\ee
where $V:\, \DZ^d\times\Om\to\DR$ is a random field on $\DZ^d$ relative to a probability space
$(\Om,\fF,\DP)$ with IID (independent and identically distributed) values.

Unlike all earlier mathematical works on localization in multi-particle models in presence of a random external
potential, we do not assume any regularity of the random amplitudes of the site potentials. The prototypical example is
given by the Bernoulli-Anderson model, but our techniques apply to arbitrary compactly supported probability
measures not concentrated on a single point.

We always assume $d\ge 2$, since the analysis of one-dimensional models calls for more optimal,
specifically one-dimensional techniques.

As was shown in \cite{CS09b}, an extension of the proof of localization for interactive $2$-particle models to
an arbitrary (but fixed from the beginning) number of particles $N$ is conceptually not difficult, except for
the proof of eigenvalue comparison estimates for norm-distant pairs of $N$-particle cubes. As the matter of fact,
the transition from $N=2$ to $N\ge 3$ aiming to prove localization estimates in the physically natural, norm-distance metric
in the $N$-particle configuration space, requires new ideas and techniques. Such a program has been carried out
in the general framework of the Multi-Scale Analysis (MSA) based on estimates in probability, as in the pioneering works
on the single-particle MSA \cite{FS83,FMSS85,DK89}; see \cite{CS17,C16b}. This task is yet to be performed in the context of the
adaptation of the Fractional Moments Method to the $N$-particle models developed by Aizenman and Warzel \cite{AW09}
(for the lattice models) and by Fauser and Warzel \cite{FW15} (in a Euclidean space).

Generally speaking, infinite-range particle-media interaction potentials make more difficult the localization analysis, and so
do singular (e.g., Bernoulli) probability measures of the random amplitudes of the site potentials. Curiously enough, a combination
of the two difficulties solves a thorny problem encountered in the eigenvalue comparison analysis of $N$-particle Anderson
Hamiltonians, particularly in a continuum configuration space. In the present paper we study a "toy-model" with piecewise-constant
("staircase") site potentials, for which the proof of eigenvalue comparison estimates is simpler than for more realistic
potentials, e.g., for $\fu(r) = r^{-A}$, $A>d$. However, it was shown in \cite{C16e} that satisfactory EV comparison estimates
can be obtained for the realistic potentials, too. The bottom line is that neither the restriction $N=2$ nor the use of the
"staircase" site potentials $\fu$ is crucial for the EV comparison estimates which, in turn, are vital for the
efficient $N$-particle localization bounds, $N\ge 2$.

Let be given a function $\DN\ni r \mapsto \fu(r)$ the function
$$
\DZ^d \ni x \mapsto \fu(|x|)
$$
is absolutely summable. Then one can define
a linear transformation $\BU$, well-defined on any bounded function $\Bfq:\, \DZ^d \to \DR$:
\be
\label{eq:def.BU}
\BU:\, \Bfq \mapsto \BU[\Bfq] \,, \;\; \text{ with } \;\;  \BU[\Bfq]:\,\DZ^d \to \DR \,,
\ee
where, setting $\rrq_y \equiv \rrq(y)$, one has
\be\label{eq:def.BU.2}
(\BU[\Bfq])(x) = \sum_{y\in\DZ^d} \fu(y-x) \fq_y .
\ee

To clarify the main ideas of \cite{C16a} and simplify some technical aspects,
the interaction potential $\fu:\, \DR_+ \to \DR$ is assumed to have the following form.
Given a real number $\kap>1$, introduce a growing integer sequence
\be
\label{def:fr.kap}
\fr_k = \lfloor k^\kap \rfloor\,, \;\; k\in\DN\,,
\ee
and let
\be
\label{eq:def.fu}
\fu(r) = \sum_{k=1}^{\infty} \fr_k^{-A} \one_{[\fr_{k}, \fr_{k+1})}(r)  \,,
\ee
Making $\fu(\cdot)$ piecewise constant will allow us to achieve,
albeit in a somewhat artificial setting, an elementary derivation of infinite smoothness of the DoS
from a similar property of single-site probability distributions of the potential $V$. We refer to $V$
as the \emph{cumulative} potential in order to distinguish it from the \emph{interaction} potential $\fu$
(which is a functional characteristics of the model) and from the local potential \emph{amplitudes}
$\{\fq_y, \, y\in\DZ^d \}$. The notation $\fq_y$ will be used in formulae and arguments
pertaining to general functional aspects of the model, while in the situation where the latter amplitudes are random
we denote them by $\om_y$.

We always assume the amplitudes $\fq_y$ and $\om_y$ to be uniformly bounded. In the case of random amplitudes,
one should either to assume this a.s. (almost surely, i.e., with probability one) or to construct from the beginning
a product measure on $[0, 1]^{\DZ^d}$ rather than on $\DR^{\DZ^d}$ and work with samples $\om\in [0, 1]^{\DZ^d}$,
which are thus automatically bounded. It is worth mentioning that boundedness is not crucial to most of the key
properties established here, but results in a streamlined and more transparent presentation. On the other hand,
as pointed out in \cite{C16e}, there are interesting models with unbounded  amplitudes $\om_\bullet$
such that $\esm{ \left(\om_\bullet - \esm{\om_\bullet}\right)^2} <\infty$. It is readily seen that
single-site probability distribution of the cumulative potential $V(x,\om)$, necessarily compactly supported
when $\om_\bullet$ are uniformly bounded and the series \eqref{eq:def.BU.2} (with $\fq_y$ replaced with
$\om_y$) converges absolutely, cannot have an analytic density, for it would be compactly supported and not
identically zero, which is impossible. However, in some class of marginal measures of $\om_\bullet$
with unbounded support, considered long ago by Wintner \cite{Wint1934} in the framework of Fourier analysis of
infinite convolutions of singular probability measures, the single-site density of $V(\cdot,\om)$ can be analytic on $\DR$.

\vskip2mm
We also always assume that  $\om_y$ are IID.

\section{Main results}
\label{sec:main.results}

\subsection{Infinite smoothness of single-site distributions}

\btm[Cf. \lcite{C16f}{Theorem 1}]
\label{thm:infinite.smooth.V}
Consider the potential $\fu(r)$ of the form \eqref{eq:def.fu}, with $A>d$ and let $d \ge 1$.
Then the characteristic functions of the random variables $V(x,\om)$ of the form \eqref{eq:def.BU.2}
obey the upper bound
$$
\big| \ffi_{V_x}(t)\big| \le \Const \eu^{ - c |t|^{d/A}} .
$$
Consequently, for any $d \ge 1$ the r.v. $V_x$ have probability  densities
$\rho_x \in \mcC^\infty(\DR)$.
\etm

\subsection{Infinite smoothness of the DoS}

\btm[Cf. \lcite{C16f}{Theorem 2}]
\label{thm:DoS.xi.Lam}
Fix a $2$-particle cube $\bball = \bball_L(\Bu)$.
\begin{enumerate}[\rm(A)]
  \item There exists a $\sigma$-algebra $\fB_\bball$,
  an $\fB_\Lam$-measurable self-adjoint random operator $\tBH_\bball(\om)$ acting in $\ell^2(\bball)$, and a
  $\fB_\bball$-independent real-valued r.v. $\xi_\bball$ such that
\be
\label{eq:H.H0.xi}
\BH_\bball(\om) = \tBH_\bball(\om) + \xi_\bball(\om) \one_\bball \,.
\ee

  \item The characteristic function $\ffi_{\xi_\bball}$ of \,$\xi_\bball$ \, satisfies the decay bound
\be
 \big|  \ffi_{\xi_\bball}(t) \big| \lea  \eu^{-|t|^{d/A}} \,.
\ee
\end{enumerate}

\etm

\subsection{Wegner estimate}

The next result is a development of \lcite{C16f}{Theorem 3} for $N$-particle Hamiltonians.

\btm["Frozen bath" Wegner estimate]
\label{thm:Wegner}
Fix the exponent $\kap>1$ in the definition \eqref{def:fr.kap} of the sequence $\fr_k$\,,
and consider a ball $\bball=\bball_L(\Bu)$ and the Hamiltonian $\BH_\bball$.
Let
\be
\label{eq:def.btau.kap}
\btaukap = \frac{\kap}{\kap-1}\,.
\ee
Fix any $\ttau>\btaukap$, consider a larger set
$\obball = \bball_{L^{\ttau}}(\Bu)\setminus\bball_{L}(\Bu)$,
and denote by $\prsub{\!\!\obball^\perp}{\cdot}$ the conditional probability given the \sigal $\fF_{\DZ^d\minus\Pi \obball}$.
Then for any $\eps \ge \eps_L := L^{- A\ttau }$
\be
\label{thm:Wegner.prob.frozen}
\prsub{\!\!\obball^\perp}{ \dist(\Sigma_{\bball}, E) \lea \eps_L }
\lea  | \bball | \, L^{\left(A-\frac{d}{2}\right){\tiny\btaukap}} \eps_L \,.
\ee
In particular, with $\ttau > 2\btau$ one has for some $c>0$
\be\prsub{\!\!\obball^\perp}{ \dist(\Sigma_{\bball}, E) \le \eps_L }\lea | \bball | \,L^{- \frac{(A+c)\ttau}{2} }.
\ee
\etm

\bco[Stable Wegner estimate]
\label{cor:Wegner.stable}
\be
\label{eq:cor.Wegner.bound}
\pr{\om_{\obball}:\, \inf_{\om=(\om_\obball, \om_\obball^\perp)} \;\; \dist\big(\Sigma(\BH_{\bball}(\om), E\big)
\lea 2\eps_L }
\lea | \bball | \, L^{- \frac{(A+c)\ttau}{2} }.
\ee
\eco

\bre
\label{rem:beta}
It follows from the construction of the event in the LHS of \eqref{eq:cor.Wegner.bound} that it is measurable
with respect to the \sigal generated by $\set{V(y,\om), \, y\in \Pi\obball}$. Consequently, any number of such events
with pairwise distant centers are independent. This observation will be useful for the proof of Lemma \ref{lem:factor.prob.S.polynom}.
\ere

\btm[Eigenvalue comparison estimate]
\label{thm:EVComp}
Fix the exponent $\kap>1$ in the definition \eqref{def:fr.kap} of the sequence $\fr_k$\,,
and consider a ball $\bball=\bball_L(\Bu)$ and the Hamiltonian $\BH_\bball$.
Then
\be
\label{thm:Wegner.prob.frozen}
\all \eps>0 \quad
\pr{ \dist\left(\Sigma\big(\BH_{\bball'}\big),\, \Sigma\big(\BH_{\bball''}\big)\right) \le \eps }
\le C \, | \bball | \, L^{\left(A - \frac{d}{2}\right){\tiny\btaukap} } \eps \,.
\ee
\etm

\subsection{Localization}

Below we denote by $\cscB_1(\DR)$ the set of all bounded Borel functions $\phi:\, \DR\to\DC$
with $\| \phi \|_\infty\le 1$. As usual, $\lr{\Bx}$ stands for $\pa{|\Bx|^2+1}^{1/2}$.

\btm[Localization at low energy]
\label{thm:main.polynom.low.E}
Consider the potential $\fu(r)$ of the form \eqref{eq:def.fu}, with $A>d>1$, and let $\fb>d/2$.
There exist an energy interval $I = [E_0, E_0 + \eta]$, $\eta>0$, near the a.s. lower edge of spectrum
$E_0$ of the random operator $\BH(\om) = -\BDelta + \BV(\Bx,\om)+\BU(\Bx)$
such that
\begin{enumerate}[\rm(A)]
  \item with probability one, $\BH(\om)$ has in $I$ pure point spectrum with square-summable eigenfunctions
  $\BPsi(\Bx,\om)$ satisfying
\be
\big| \BPsi(\Bx,\om) \big| \le C_{\BPsi}(\om) \, \lr{\Bx}^{-\fb};
\ee

  \item for any $\Bx,\By\in\DZ^d$ and any connected subgraph $\mcG \subseteq \DZ^d$  containing $x$ and $y$ one has
\be
\esm{ \sup_{\phi \in\cscB_1(\DR)}
\left\| \one_\Bx \, \rP_I\big(\BH_\mcG(\om)\big) \phi\big(\BH_\mcG(\om)\big)  \,\one_\By  \right\| }
   \le \lr{|\Bx-\By|}^{-\fb} .
\ee
\end{enumerate}
\etm

In the present paper, we focus mainly on the strong dynamical localization and privilege clarity of constructions
and proofs, and for these reasons we use the fixed-energy MSA induction: it is substantially simpler than its
variable-energy (energy-interval) counterpart initially developed in \cite{CS09b,CBS11} and streamlined by Klein and Nguyen
\cite{KN13,KN16}. As it is well-known by now, the energy-interval MSA estimates, crucial to the proofs of spectral and dynamical
localization, can actually be inferred from the fixed-energy variants without actually carrying out a separate energy-interval
scale induction (the latter path has been employed by Germinet and Klein in \cite{GK01} and subsequent papers). However,
the direct derivations require some additional information and arguments, and the efficiency of the final estimates
depends on the strength of the fixed-energy probabilistic bounds (cf. \cite{ETV10}) and on the specific form of the
IAD (Independence At Distance) property featured by the model (cf. \cite{BK05}).

Speaking of the method proposed by
Elgart \etal \cite{ETV10}, improving an older observation made by Martinelli and Scoppola \cite{MS85},
an \emph{exponential} decay of the eigenfunctions in the localization interval of energies requires
exponentially decaying probability bounds on unwanted events in the course of the fixed-energy analysis,
and such bounds cannot be achieved today by the existing MSA techniques.

An alternative to the method of \cite{ETV10} was proposed in \cite{C14a}, and in the context of $N$-particle
Anderson Hamiltonians in a Euclidean space it was used in \cite{CS17} for the analysis of $N$-particle lattice models,
and in \cite{C16b} where a particular class of alloy potentials
("flat tiling" potentials) was studied. The specificity of the "flat tiling" alloys is that the sample space
contains piecewise-constant functions whose plateaus can cover arbitrarily large cubes. A thorough analysis of the
"staircase" potentials $\fukap$ considered in the present paper shows that one can use a similar (and actually, even a
slightly simpler) technique, and thus prove exponential spectral localization. I plan to provide the details in a forthcoming
work.

The reader can also see that the main ingredients required for the energy-interval $N$-particle MSA induction
are obtained in Sections \ref{sec:Wegner} (eigenvalue concentration estimates for individual cubes) and
\ref{sec:EV.compare.staircase} (eigenvalue comparison estimates for pairs of cubes). Therefore, a more tedious,
direct proof of energy-interval estimates, leading to the exponential spectral localization, can also be obtained.

\section{Fourier analysis of probability measures}
\label{sec:char.f.polynom}

\subsection{The Main Lemma}

\ble[Cf. \lcite{C16e}{Lemma 4.1}]
\label{lem:Main}
Let be given a family of IID r.v.
$$
X_{n,k}(\om), \; n\in \DN, \;\; 1 \le k \le K_n \,, \;\; K_n \asymp n^{d-1},
$$
and assume that their common characteristic function $\ffi_X(t) = \esm{ \eu^{\ii t X}}$ fulfills
\be
\label{eq:Main.Lemma.ffi.t0}
\ln \, \big| \ffi_X(t) \big|^{-1} \ge C_X t^2, \;\; |t|\le t_0.
\ee
Let
$$
\bal
S(\om) &= \sum_{n\ge 1} \sum_{k=1}^{K_n} \fa_n X_{n,k}(\om), \;\; \fa_n \asymp n^{-A} \,,
\\
S_{M,N}(\om) &= \sum_{n=M}^N \sum_{k=1}^{K_n} \fa_n X_{n,k}(\om), \;\; M \le N \,.
\eal
$$
The the following holds true.
\begin{enumerate}[\rm(A)]
  \item\label{item:Main.Lemma.A}
There exists $C = C(C_X, t_0, A,d)\in(0,+\infty)$ such that
$$
\all t\in\DR  \quad \big| \ffi_S(t) \big| \le C \eu^{- |t|^{d/A}} \,.
$$

  \item\label{item:Main.Lemma.B}
   For any  $N \ge (1+c)M \ge 1$ with $c>0$,  and $t$ with $|t| \le N^{A}$,
$$
\ln \left| \esm{ \eu^{\ii t S_{M,N}(\om)}} \right|^{-1}
\gea  M^{-2A+d} \,t^2 \,.
$$

  \item\label{item:Main.Lemma.C}
  Let $I_\eps = [a,a+\eps]\subset \DR$. Then for any $\eps \ge N^{-A}$ one has
\be
\label{eq:Main.Lemma.D}
\pr{ S_{M,N}(\om) \in I_\eps } \lea  M^{A - \frac{d}{2} } \, \eps\,,
\ee
\end{enumerate}
\ele

\subsection{Thermal bath estimate for the cumulative potential}
\label{ssec:thermal.bath.ch.f}

Here we recall some of the results obtained in \cite{C16d}.
\ble
\label{lem:thermal.bath.F.V}
Consider a random field $V(x,\om)$ on $\DZ^d$ of the form
$$
V(x,\om) = \sum_{y\in\DZ^d} \fu(y-x)\, \om_y \,,
$$
where $\fu$ is given by \eqref{eq:def.fu} and $\{\om_x, \, x\in\DZ^d\}$ are bounded IID r.v. with nonzero variance. Then the following holds true:

\begin{enumerate}[\rm(A)]
  \item The common characteristic function $\ffi_V(\cdot)$ of the identically distributed r.v.
$V(x,\om)$, $x\in\DZ^d$, obeys
\label{lem:thermal.bath.bound}
\be
\all t\in\DR \quad \big| \ffi_V(t)\big| \le C \, \eu^{-|t|^{d/A}} \,.
\ee

  \item Consequently, the common probability distribution function $F_V(\cdot)$ of the
  cumulative potential at sites $x\in\DZ^d$ has the derivative $\rho_V \in\mcC(\DR)$.

  \item Let $v_* := \inf\, \supp \rho_V$, then $F_V(v_*+\lam) = \ord{|\lam|^\infty}$.
\end{enumerate}

\ele

\section{Infinite smoothness of the DoS and Wegner estimates}
\label{sec:Wegner}

\subsection{DoS in a thermal bath}
\label{ssec:inf.smooth.DoS.thermal.bath}

\proof[Proof of Theorem \ref{thm:DoS.xi.Lam}] \textbf{Assertion (A)}.
Fix a $2$-particle cube $\bball_L(\Bu)$, $\Bu = (u_1, u_2)$, and consider the two possible situations.
\par\vskip1mm\noindent
\textbf{Case (I)} $|u_1 - u_2| \le 4L$.

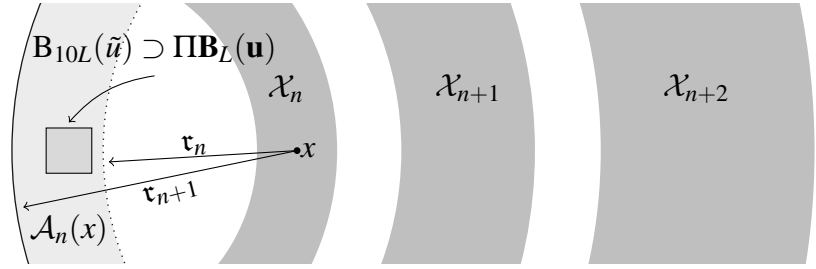
\begin{figure}
\begin{tabular}{c}
%
%
%
\begin{tikzpicture}
\begin{scope}[scale=0.15]
\begin{scope}
\clip (-15,-12) rectangle ++(100.0,28.0);

\fill[color=white] (0, -11) circle (0.2);
\fill[color=white] (0, 14) circle (0.2);

\clip (-10,-10) rectangle ++(100.0,23.0);

\fill[color=white] (-11,0) circle (0.2);

\draw[color=white!85!gray, line width = 34] (30, 0) circle (31);

\fill[color=white!70!gray] (-2, -2) rectangle ++(4, 4);
\draw (-2, -2) rectangle ++(4, 4);

\node[](An) at (0, -7) {$\mcA_n(x)$};

\begin{scope}
\draw[color=white!50!gray, line width = 30] (0, 0) circle (20);
\end{scope}

\draw[dotted, color=white!10!black, line width = 0.5] (30, 0) circle (27);
\draw[color=white!10!black, line width = 0.5] (30, 0) circle (35);

\fill (33, 17.5) circle (0.34);
\draw[->] (20, 0) -- (3.5,-1);
\draw[->] (20, 0) -- (-4, -5);

\node[rotate=5](r2) at (11, 0.5) {$\fr_{n}$};
\node[rotate=14](r1) at (9, -3.7) {$\fr_{n+1}$};

\node (cXn) at (19, 5.5) {$\mcX_n$};

\node[](x) at (21, 0) {$x$};
\fill(20,0) circle (0.3);

\begin{scope}
\clip (20,-10) rectangle ++(100.0,23.0);
\draw[color=white!50!gray, line width = 50] (0, 0) circle (35);

\node (cXnplus) at (35, 5.5) {$\mcX_{n+1}$};

\draw[color=white!50!gray, line width = 80] (0, 0) circle (56);
\node (cXnplustwo) at (55, 5.5) {$\mcX_{n+2}$};

\end{scope}

\node (ball) at (7.5, 9.0) {$\ball_{10L}(\tu)\supset \Pi\bball_L(\Bu)$};
\draw[->,bend right = 20] (ball.south) to (0, 2.5);

\end{scope}

\end{scope}

\end{tikzpicture}

\end{tabular}

\caption{  \footnotesize
Here is shown the physical, \textbf{single-particle}
space $\mcZ \equiv \DZ^d$ and not the multi-particle, product space.
For each fixed $x\in\mcX_{n}$, $n\ge n_\circ(L,\varkappa)$, the potential $\om_x \fukap(x-\cdot)$
takes a constant value on an annulus
$\mcA_n(x)=\ball_{\fr_{n+1}}(x)\moins\ball_{\fr_{n}}(x)$ (the leftmost light-gray arc),
hence on the entire cube $\ball_L(0)\subset\mcA_n(x)$. Therefore the sum of such potentials
is a random constant on $\ball_L(0)$ with a smooth probability measure. The regularity of the latter
can be assessed essentially in the same way as for the individual values of the cumulative potential
$V(y,\om)$, $y\in\ball_L(0)$. The remaining potentials $\om_x \fukap(x-\cdot)$
(those which are non-constant on $\ball_L(0)$) can be rendered non-random
by conditioning.}
\label{lab:fig01}
\end{figure}%
%

\vskip1mm
Then $\exist \tu\in\DZ^d$ such that (see Fig.~\ref{lab:fig01})
$$
\Pi \bball_L(\Bu) \equiv \ball_L(u_1) \cup \ball_L(u_2)\subset\tball:=\ball_{10 L}(\tu).
$$
In this case, we can argue as in \cite{C16f} and find an infinite set of sites $\mcX(\tball)$ such that
the random function on the lattice
$$
(x,\om) \mapsto \sum_{y\in \mcX(\tball)} \om_y\fu(|y-x|)
$$
generates on $\tball$ (hence on both 1-particle projection cubes $\ball_L(u_i)$, $i=1,2$) a random,
constant in space potential, viz.
$$
(x,\om) \mapsto \sum_{y\in \mcX(\tball)} \om_y\fu(|y-\tu|)\one_{\tball}(x) = \xi(\om)\one_{\tball}(x),
$$
where the r.v. $\xi$ has an infinitely smooth probability measure.

For the $2$-particle potential generated on $\bball$ this gives $2\xi(\om)\one_{\Pi \bball}(x)$,
since both projections of $\bball$ are affected by the same random constant.

Obviously, all the EVs of the Hamiltonian
in $\bball_L(\Bu)$ subject only to the above random potential admit a representation
$$
E_i(\om) = \lam_i + \xi(\om),
$$
with non-random shifts $\lam_i$, whence the infinite smoothness of their individual probability measures.

The total random potential induced on $\bball_L(\Bu)$ is decomposed into the sum
$$
\BV_{\bball}(\Bx,\om) = \BW_{\bball}(\Bx,\om) + 2\xi(\om) \one_{\bball}(\Bx),
$$
where $\BW_{\bball}(\Bx,\om)$ is independent of $\xi(\om)$ since it is generated by the random amplitudes
nt encountered in $\xi(\om)$. Therefore, we can first condition on $\BW_{\bball}(\Bx,\om)$ and obtain
an infinitely smooth probability measure for each random EV
$$
\om \mapsto E_i(\om) = \lam_i(\om) + 2\xi(\om)
$$
with the shift $\lam_i(\om)$ rendered nonrandom by conditioning,
and then switch $\BW_{\bball}(\Bx,\om)$ on,
thus obtaining the a priory (unconditional) probability measure of $E_i(\cdot)$ as
the convolution of the two independent random summands $\lam_i(\om)$ and $2\xi(\om)$.
The resulting convolution measure is at least as smooth as the one of $2\xi(\om)$.

\vskip7mm
\par\vskip1mm\noindent
\textbf{Case (II)} $|u_1 - u_2| > 4L$.

\begin{figure}
\begin{tabular}{c}
%
%
%
\begin{tikzpicture}

\begin{scope}[scale=0.15]
\clip (-17,-9) rectangle ++(70.0, 32.0);

\fill[color=white] (0, -9) circle (0.2);
\fill[color=white] (0, 14) circle (0.2);

\begin{scope}
\clip (-17,-7) rectangle ++(70.0,28.0);

\fill[color=gray!20!white]  (12, 2) rectangle ++(4, 4);
\draw  (12, 2) rectangle ++(4, 4);
\fill (14,4) circle (0.2);

\fill[color=gray!20!white] (-2, -2) rectangle ++(4, 4);
\draw (-2, -2) rectangle ++(4, 4);
\fill (0,0) circle (0.2);

\node (ball) at (-8, 4.7) {$\ball_L(u_2)$};
\draw[->,bend left = 40] (ball.east) to (0.2, 2.4);

\node (ball) at (5.0, 9.0) {$\ball_L(u_1)$};
\draw[->,bend left = 40] (ball.east) to (14.0, 6.3);

\node (cX) at (32, 3.0) {$\mcX_n$};
\draw[->,bend left = 20] (cX.north) to (36, 7);
\begin{scope}
\clip (5,-20) rectangle ++(20.0, 40.0);
\draw (33, 17.5) circle (20);
\end{scope}

\draw[line width=2.0] (33, 17.5) circle (34);

\begin{scope}
\draw[color=white!50!gray, line width = 8] (37, 0) arc (-14:46:20);
\end{scope}

\draw[dotted] (0,0) -- (14,4);
\draw (14,4) -- (49,14);

\fill (33, 17.5) circle (0.34);
\draw[->] (33, 17.5) -- (13.9,12.5);
\draw[->] (33, 17.5) -- (-0.5,14);

\node[rotate=5](r2) at (20, 17.5) {$\fr_{n+1}$};
\node[rotate=14](r1) at (22, 13.2) {$\fr_{n+1} - CL$};

\draw[<->] (3.0,-1.29) -- (11,1.2);
\node[rotate=17](R) at (8.5, -1.7) {$C' L$};

\draw[->, dotted] (14,4) -- (33, 17.5);
\node (C) at (24.2, 6.8) {};
\draw[->, bend right = 20] (C.north) to (23.1, 10.0);
\node[rotate=17](CT) at (25.2, 9.2) {$\theta$};
\node[rotate=40](CR) at (28, 12.8) {$r$};

\node[](x) at (37.0, 18.5) {$x\in\mcX_n$};

\end{scope}

\end{scope}

\end{tikzpicture}

\end{tabular}

\caption{  \footnotesize\emph{Example for the case {\rm\textbf{(II)}}.}
}
\label{lab:fig02}
\end{figure}
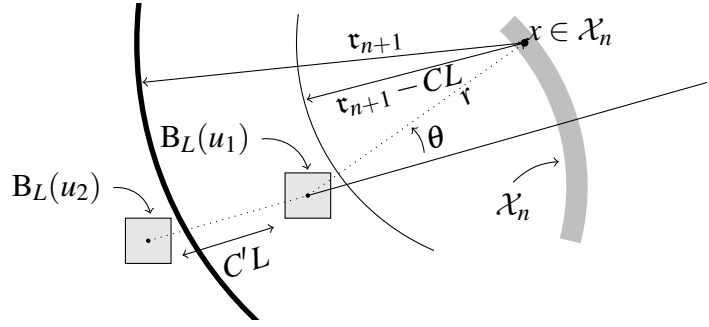

In this case we can arrange an infinite sequence of scatterers' subsets $\mcX_n$ which induce on each of the projection
cubes $\ball_L(u_1)$ and $\ball_L(u_2)$ respective constant random fields, albeit with different values of the random constants
(see Fig.~\ref{lab:fig02}),
$$
\bal
\ball_L(u_1) \ni &x \mapsto a_x \xi_x(\om) \one_{\ball_L(u_1)} ,
\\
\ball_L(u_2) \ni &x \mapsto c_x a_x \xi_x(\om) \one_{\ball_L(u_2)} ,
\eal
$$
with $c_x \in [0,1]$,
and the resulting random potential induced by the scatterer at $x$ thus acts as a random scalar operator
$\big(1 + c_x \big) a_x \xi_x(\om) \one_{\bball_L(\Bu)}$.
Since $(1+c_x)a_x \asymp a_x$, we conclude
as in case \textbf{(I)} that the convolution of all admissible r.v.
$\big(1 + c_x \big) a_x\xi_x$ has a $\mcC^\infty(\DR)$-density.

%
%


\vskip3mm
\noindent
\textbf{Assertion (B)}.
The claim follows easily from the Main Lemma \ref{lem:Main}; we only need to identify
the key ingredients  of the latter:
$$
\bal
\mcX_n &: = \myset{x\in\DZ^d:\, \dist\left(x, \Lam_L\right) \in[\fr_n, \fr_{n+1}) }\,,
\;\; K_n := \big|\mcX_n \big| \,,
\\
\myset{ \om_x, \, x\in\mcX_n } & \leftrightarrow \myset{ X_{n,k}, \, k=1, \ldots, K_n }
\\
M &:= L, \;\; N = +\infty\,,
\\
S_{M,N}(\om) &= \sum_{n=M}^\infty \sum_{k=1}^{K_n} \fa_n X_{n,k} \equiv \sum_{x:\, |x|\ge L} \fu(|x|) \om_{x}
\eal
$$
\qedhere

\subsection{Wegner estimates}
\label{ssec:Wegner}

Aiming to the applications to Anderson localization, we now have to operate
with a restricted, annular "bath" of finite size, the complement of which is "frozen".
This is necessary for obtaining a satisfactory replacement for the IAD property
very valuable in the short-range interaction models.

\proof[Proof of Theorem \ref{thm:Wegner}]
The required bound follows from assertion (C) of Lemma \ref{lem:Main}.

Identification of the principal ingredients of
Lemma \ref{lem:Main} is as follows:
$$
\bal
\mcX_n &: = \myset{x\in\DZ^d:\, |x| = n }\,, \;\; K_n := \big|\mcX_n \big| \,,
\\
\myset{ \om_x, \, x } & \leftrightarrow \myset{ X_{n,k}, \, k=1, \ldots, K_n }
\\
M &:= L^{\btaukap}, \;\; \text{ with } \btaukap \text{ defined in \eqref{eq:def.btau.kap}}
\\
N &= L^\ttau\,, \;\; \ttau > \btaukap\,,
\\
S_{M,N}(\om) &= \sum_{n=M}^N \sum_{k=1}^{K_n} \fa_n X_{n,k} \equiv \sum_{x:\, |x|\in[L, R_L]} \fu(|x|) \om_{x}
\eal
$$
Proceeding as in Theorem \ref{thm:DoS.xi.Lam}, we obtain the representation
\be
\label{eq:thm.Wegner.frozen.xi}
\BH_\bball(\om) = \tBH_\bball(\om) + \xi_\bball(\om) \, \one_\bball\,,
\ee
where the random operator $\tBH_\bball(\om)$ is independent of the r.v. $\xi_\bball(\om)$.
By Lemma \ref{lem:Main}, $\xi_\bball$ fulfills, for any interval $I$ of length
\be
\label{eq:rel.eps.L}
\eps_L = N^{-A} \equiv L^{ -A\ttau}
\ee
the concentration estimate (cf. \eqref{eq:Main.Lemma.D})
\be
\pr{ \xi_\bball(\om) \in I_{\eps_L} } \lea M^{A-\frac{d}{2}} \, \eps_L \equiv C\, L^{\left(A-\frac{d}{2}\right)\btaukap} \, \eps_L
\lea L^{ -A\ttau + \left(A-\frac{d}{2}\right)\btaukap}\,.
\ee
In particular, with $\ttau\ge 2\btaukap$ we have
\be
\pr{ \xi_\ball(\om) \in I_{\eps_L} } \lea L^{ -\half A\ttau -\frac{d\ttau}{4}}\,.
\ee
This proves the EVC estimate \eqref{eq:cor.Wegner.bound}, since $\BH_\bball(\om)$
acts in the Hilbert space $\ell^2(\bball)$ of finite dimension $|\bball|$.
\qedhere

\section{Eigenvalue comparison bound. Proof of Theorem \ref{thm:EVComp}}
\label{sec:EV.compare.staircase}

Consider two $2$-particle cubes, $\bball'=\bball_L(\Bu^1)$ and $\bball''=\bball_L(\Bu^2)$, of radius $L$,
with $|\Bu^1 - \Bu^2|> \hC L$.
Introduce the lattice subsets
$$
\mcX_n : = \myset{x\in\DZ^d:\, \dist\left(x, \ball_L(0) \right) \in[\fr_n, \fr_{n+1}) }\,.
$$
and the spherical layers $\mcA_r = \{x:\, |x|\in[r, r +1)\}$, $r\in\DN$.
We will have to work again with $\fr_n \le r < \fr_{n+1}$, where $n$ suits the conditions $\fr_{n+1} - \fr_n \ge CL$
(cf. \lcite{C16e}{Eqn (6.3)--(6.4)},
thus with $\fr_n \gg L$. Elementary geometric arguments show that if
$\dist\left(\bball_L(\Bu^1, \bball_L(\Bu^2) \right)\ge C L$, with a sufficiently large $C>0$,
then there exist constants $C_1, C_2, c>0$ depending on the dimension $d$ with the following properties.
Exchanging if necessary $\Bu^1 \leftrightarrow \Bu^2$ and then $u^1_1 \leftrightarrow u^1_2$
(cf. Fig.~\ref{lab:fig03}),
one can find an infinite sequence of lattice subsets $\mcX_n\subset\DZ^d$, $n\ge n_\circ$
such that for all $x\in\mcX_n$, with
\be
\label{eq:cond.r.theta}
\bal
|x - u^1_1| = r & \in \big[ \fr_{n+1} - C_1\, L, \, \fr_{n+1} - C_2\, L \big] \,
\\
 \big| \cos \theta \big| &\le c
\eal
\ee
we have for all $x\in\mcA_r$, with some $n_1 \ge n+1$,
\beal
\fukap(x-y) \one_{\ball_L(u^1_1)}(y) &= \fr_n^{-A} \one_{\ball_L(u^1_1)}(y)
\\
\fukap(x-y) \one_{\ball_L(u^1_2)}(y) &= \fr_{n_1}^{-A} \one_{\ball_L(u^1_2)}(y) \,,
\eeal
while for some $n_2, n_3\ge n + \fn(\hC)$, where $\fn(\hC)\to+\infty$ as $\hC\to+\infty$,
\beal
\fukap(x-y) \one_{\ball_L(u^2_1)}(y) &= \fr_{n_2}^{-A} \one_{\ball_L(u^2_1)}(y) \,,
\\
\fukap(x-y) \one_{\ball_L(u^2_2)}(y) &= \fr_{n_3}^{-A} \one_{\ball_L(u^2_2)}(y) \,.
\eeal
%

\begin{figure}
\begin{tabular}{c}
%
%
%
\begin{tikzpicture}
\begin{scope}[scale=0.15]
\clip (-25,-5) rectangle ++(70.0,26.0);
\draw (-2, -2) rectangle ++(4, 4);
\fill (0,0) circle (0.2);

\fill[color=white!80!gray] (-25, -5) rectangle ++(4, 4);
\draw (-25, -5) rectangle ++(4, 4);
\fill (14,4) circle (0.2);

\fill[color=white!80!gray] (-22, 11) rectangle ++(4, 4);
\draw (-22, 11) rectangle ++(4, 4);
\fill (14,4) circle (0.2);

\node (balltwob) at (-18, 6.0) {$\ball_L(u^2_2)$};
\draw[->,bend left = 20] (balltwob.south) to (-22, -3);

\node (balltwoa) at (-9, 11.0) {$\ball_L(u^2_1)$};
\draw[->,bend right = 20] (balltwoa.west) to (-19, 12.9);

\draw (12, 2) rectangle ++(4, 4);
\fill (14,4) circle (0.2);

\node (ball) at (-9, -2.0) {$\ball_L(u^1_2)$};
\draw[->,bend left = 20] (ball.east) to (-1.2, 1.0);

\node (ball) at (5, 9.5) {$\ball_L(u^1_1)$};
\draw[->,bend left = 20] (ball.east) to (13.2, 5.0);

\node (cX) at (32, 3.0) {$\mcX_n$};
\draw[->,bend left = 20] (cX.north) to (36, 7);
\begin{scope}
\clip (5,-20) rectangle ++(20.0, 40.0);
\draw (33, 17.5) circle (20);
\end{scope}

\draw[line width=1.5] (33, 17.5) circle (34);

\begin{scope}
\draw[color=white!50!gray, line width = 8] (37, 0) arc (-14:46:20);
\end{scope}

\draw[dotted] (0,0) -- (14,4);
\draw (14,4) -- (49,14);

\fill (33, 17.5) circle (0.34);
\draw[->] (33, 17.5) -- (13.9,12.5);
\draw[->] (33, 17.5) -- (-0.5,14);

\node[rotate=5](r2) at (20, 17.5) {$\fr_{n+1}$};
\node[rotate=14](r1) at (22, 13.2) {$\fr_{n+1} - CL$};

\draw[<->] (3.0,-1.29) -- (11,1.2);
\node[rotate=17](R) at (8.5, -1.7) {$C' L$};

\draw[->, dotted] (14,4) -- (33, 17.5);
\node (C) at (24.2, 6.8) {};
\draw[->, bend right = 20] (C.north) to (23.1, 10.0);
\node[rotate=17](CT) at (25.2, 9.2) {$\theta$};
\node[rotate=40](CR) at (28, 12.8) {$r$};

\node[](x) at (39.0, 18.5) {$x\in\mcX_n$};

\end{scope}
\end{tikzpicture}

\end{tabular}

\caption{  \footnotesize\emph{Example for Section \ref{sec:EV.compare.staircase}.} For $y\in\ball_L(u^1_1)$
one has $\fukap(x-y)= \fr_n^{-A}$, while for $y\in\ball_L(u^1_2)$,  $\fukap$ jumps to the next plateau:
$\fu^{(\varkappa)}(x-y)= \fr_{n+1}^{-A}$. The separation sphere between the two plateaus is indicated by the thick black circle.
This sphere depends of of course upon its centre $x$, but for all $x\in\mcX$ obeying \eqref{eq:cond.r.theta} with a suitable $c>0$,
the separation does occur with the same the radii $\fr_{n+1}$, $\fr_{n+1} - CL$.
}
\label{lab:fig03}
\end{figure}%
%


Fix a measurable labeling of the eigenvalues of $\BH' = \BH_{\bball'}(\om)$ and
$\BH'' = \BH_{\bball''}(\om)$, in increasing order: $\{\lam'_\ra, \ra\in\llb 1, |\bball'| \rrb \}$ and, respectively,
$\{\lam''_\rb, \rb\in\llb 1, |\bball''| \rrb\}$. Denote
\be
\xi_{\ra, \rb}(\om) =  \lam'_{\ra}(\om) -  \lam''_{\rb}(\om)
\ee
and for the rest of the argument, fix some pair of indices
$(\ra, \rb) \in \llb 1, |\bball'| \rrb  \times \llb 1, |\bball''| \rrb $.

Let $\mcX = \cup_{n\ge n_0(x''} \mcX_n$, and decompose $\om = (\om_\mcX, \om_{\mcX}^\perp)$. From this point on,
$\om_\mcX^\perp$ will be fixed, so the probabilistic estimates will be made with respect to the conditional probability
$\pr{\cdot \cond \fF_\mcX^\perp}$.
Writing
$$
V(y,\om) = \tW(\om_\mcX^\perp) + W(\om_\mcX)(y), \;\;
W(\om_\mcX)(y) = \sum_{n \ge n_0} \sum_{x\in\mcX_n} \om_x \fukap(x-y) \,,
$$
one obtains by straightforward calculations that
\be
\xi_{\ra,\rb}(\om) = c_{\ra,\rb}(\om_\mcX^\perp) + \eta_{\ra,\rb}(\om_\mcX) \,,
\ee
where
$\eta_{\ra,\rb}$ has a $\cC^\infty$-density $p_{\ra,\rb}(\cdot)$ with
$$
\bal
\|p_{\ra,\rb}(\cdot)\|_\infty
&
\lea L^{c_\kap A}, \;\; c_\kap = \frac{\kap}{\kap-1}.
\eal
$$
For example, $c_\kap \le 2$ for $\kap\ge 2$.

Now the claim follows by counting the number of pairs $(\ra,\rb)$, which is $\Ord{L^{2d}}$.
\qed

\section{ILS estimates at low energies via "thin tails"}
\label{sec:ILS}

\btm[Stable ILS estimate]
Fix any $L_0>1$ and consider the Hamiltonian $H_{\ball_{L_0}(u)}(\om)$ with an arbitrary $u\in\DZ^d$.
Assume that the interaction potential is positive and decays as $\fu(r) = r^{-A}$, $A>d$, and introduce a larger ball
$\ball^+=\ball_{\fc L_0}(u)$ and the sigma-algebras
\begin{itemize}
  \item $\fF_{\ball^+}$ generated by all scatterers' amplitudes $\om_y$ with $y\in \ball^+$,

  \item $\fF_{\ball^+}^\perp$ generated by all scatterers' amplitudes $\om_y$
  with $y\in\DZ^d\setminus \ball^+$.
\end{itemize}
Then for any $\theta\in(0,1)$ there exists some $C_\theta>0$ such that
\be
\pr{ \om_{\Pi\bball_L(\Bu)}:\, \; \inf_{\om = (\om_{\Pi\bball_L(\Bu)}, \om_{\Pi\bball_L(\Bu)}^\perp)}
   \;\;  E_0^\Lam(\om) \le L_0^{-\theta} } \le  \eu^{ - C_\theta L_0^d } \,.
\ee
Consequently, for any nontrivial compactly supported probability measure of the random amplitudes $\om_\bullet$
of the site potentials $x \mapsto \fu(|\bullet - x|)$, for any $b>0$ there exists a nontrivial interval $I_* = [0,E_*]$
and $L_0$ large enough such that one has
$$
\pr{\om_{\Pi\bball_L(\Bu)}:\, \; \inf_{\om = (\om_{\Pi\bball_L(\Bu)}, \om_{\Pi\bball_L(\Bu)}^\perp)}
   \;\;   E_0^\Lam(\om) \le E_* } \le L_0^{-b}.
$$

\etm

\proof
Due to the assumed positivity of the interaction potential $U^{(2)}$,
we have with $\bball = \ball'\times\ball''$
\beal
\BH_{\bball_L(\Bu)}(\om) & \ge  H_{\ball_L(u_1)}(\om)\otimes \one_{\ball_L(u_2)}
+ \one_{\ball_L(u_1)} \otimes  H_{\ball_L(u_2)}(\om) .
\eeal
It has been noticed already in earlier works on multi-particle Anderson Hamiltonians (with regular probability
distribution of the amplitudes of the site potentials) that the ILS estimate for $N$-particle Hamiltonians
follow directly for their $1$-particle counterparts (projection Hamiltonians). The derivation itself
does not rely on the regularity properties of the disorder distribution, so the claim actually follows from
its $1$-particle variant established in \cite{C16e}.
\qedhere

\section{Proof of localization}
\label{sec:loc.polynom}

\subsection{Deterministic analysis}

We adapt the strategy from \cite{KirStoStolz98a}.

Working with a Hamiltonian $H_{\bball_L(u)} = -\Delta_{\bball_L(u)} + gV$
in a given cube $\bball_L(u)$, it will be necessary to know the values of the amplitudes $\om_y$
with $y$ in a larger cube $\bball_{R_L}(u)\supset\bball_L(u)$, where the specific choice of $R_L$
depends upon the decay rate $r \mapsto r^{-A}$ of the interaction potential $\fu(r)$, along with
some other parameters of the model and of the desired rate of decay of EFCs to be proved.
Below we set $R_L = L^\ttau$, $\ttau>1$.

\bde
A $2$-particle cube $\bball_L(\Bu) = \ball_L(u_1)\times \ball_L(u_2)$ is called non-interactive (NI) if
$|u_1 - u_2|> 4 L$, and partially interactive (PI), otherwise.
\ede

\bde
\label{def:NS.NR.polynom}
Let be given a cube $\bball=\bball_L(u)$.
A configuration $\Bfq\in \BfQ_{\DZ^d}$
is called
\begin{enumerate}[\rm(1)]
  \item $(E,\delta,\bball)$-non-singular iff the resolvent $G_{\bball}(E)$ of the operator
$$
H_{\bball_L(u)} = -\Delta_{\bball_L(u)} + \BU[\Bfq]\big|_{\bball_{L}}
$$
  (cf. the definition of $\BU[\Bfq]$ in \eqref{eq:def.BU.2}) is well-defined and satisfies
\be
\label{eq:def.NS}
\max_{x\in \bball_{L/3}(u)} \; \max_{y\in \pt^-\bball_{L}(u)}
 \; \left\| G_{\bball}(x,y; E) \right\| \le \delta\,;
\ee

  \item $(E,\eps,\bball)$-non-resonant iff
\be
\label{eq:def.NR}
  \dist\big( \Sigma( H_{\bball} ), \, E \big) \ge \eps.
\ee

\end{enumerate}
\ede

When the condition \eqref{eq:def.NS} (resp., \eqref{eq:def.NR}) is violated,
$\Bfq$ will be called $(E,\eps)$-singular (resp., $(E,\eps)$-resonant).
We will be using obvious shortcuts $(E,\eps)$-NS, $(E,\del)$-S, $(E,\eps)$-NR, $(E,\eps)$-R.

\bde
\label{def:SNS.SNR.polynom}
Let be given a cube $\bball=\bball_L(\Bx)$ and a real number $\ttau>1$. Denote $\obball=\bball_{L^\ttau}(\Bx)$.
A configuration $\Bfq_{\Pi\bball_{L^\ttau}(\Bx)}\in\BfQ_{\Pi\bball_{L^\ttau}(\Bx)}$
is called
\begin{enumerate}
  \item $(E,\eps)$-SNS (strongly non-singular in $\bball$, or stable non-singular) iff
   for any configuration of amplitudes $\Bfq_{(\Pi\obball)^\rc}\in \BfQ_{(\Pi\obball)^\rc}$
   the extension of $\Bfq_{\Pi\obball}$ to the entire lattice,
   $\Bfq = (\Bfq_{\Pi\obball},\Bfq_{(\Pi\obball)^\rc})$ is $(E,\eps)$-NS
   in $\bball$;

  \item $(E,\eps)$-SNR (strongly NR, or stable NR)
   iff for the zero-configuration  $\BfQ_{(\Pi\obball)^\rc} \ni \Bfq_{(\Pi\obball)^\rc} \equiv 0$
   the function
   $V_\bball = \BU[\Bfq_{\obball} + \Bfq_{\obball^\rc}]\big|_{\obball}
    = \BU[\Bfq_{\obball}]\big|_{\obball}$ is
   $(E, \eps)$-CNR.

   \item $(E,\delta,K)$-strongly-good ($(E,\delta,K)$-S-good) in $\bball$ iff for the configuration  $\BfQ_{(\Pi\obball}$
   the cube $\bball$ contains no collection of $K$ or more PI cubes $\bball_{L_k}(\Bu^i)$,
   with pairwise $L^\ttau$-distant centers $\Bu^i$, none of which is $(E,\delta)$-SNS.
\end{enumerate}
\ede

In subsection \ref{ssec:scaling} we work in the situation where the potential $V:\DZ^d\to\DR$ is fixed,
and perform a deterministic analysis of finite-volume Hamiltonians.
It will be convenient to use a slightly
abusive but fairly traditional terminology and attribute the non-singularity and non-resonance
properties to various cubes $\bball$ rather than to a configuration $\Bfq$ or a cumulative potential $V =
\BU[\Bfq]$,
which will be fixed anyway. Therefore, we will refer, for example, to $(E,\eps)$-NS balls
instead of $(E,\eps,\bball)$-NS  configurations $\Bfq$.

\subsection{Scaling scheme}
\label{ssec:scaling}
Fix $\DN\ni d\ge 1$, $A>d$ and the interaction potential $\fu(r) \;\big(\sim r^{-A} \, \big)$ of the form
\eqref{eq:def.fu}.
Further, fix an arbitrary number $b>d$, which will represents the desired polynomial decay rate of the key
probabilities in the MSA induction, and let
\begin{align}
\label{eq:con.alpha.tau.S}
\alpha & >  \ttau > \frac{b}{A-d } \,, \quad
\DN \ni S > \frac{b\alpha}{b - \alpha d} \,, \quad
L_{k+1} = \big\lfloor L_k^{\alpha} \big\rfloor \,,  \;\; k\ge 0 \,,
\end{align}
with $L_0$ large enough, to be specified on the as-needed basis. A direct analog of the well-known
deterministic statement \lcite{DK89}{Lemma 4.2} is the following statement adapted to long-range interactions
essentially as in \cite{KirStoStolz98a}. Denote

\be
\label{eq:def.eps.k.AL}
m_k := \left(1 + L_k^{-1/8}\right)m \,, \;\;
\eps_k := 4 L_k^{-\left(A - \frac{d}{2}\right)\ttau}\,,\;\
\delta_k := \eu^{-m_k L_k} \,.
\ee

\ble[Conditions for strong non-singularity]
\label{lem:cond.SNS.polynom}
Consider a cube $\bball=\bball_{L_{k+1}}(\Bu)$, $k\ge 0$, and suppose that
\vskip1mm
\par\noindent
{\textnormal{\textbf{(i)}}} $\bball$ is $(E, \eps_k)$-SNR;
\vskip1mm
\par\noindent
{\textnormal{\textbf{(ii)}}} all non-interactive cubes $\bball_{L_k}(\Bx)\subset\bball$
are $(E, \delta_k)$-SNS;
\vskip1mm
\par\noindent
{\textnormal{\textbf{(iii)}}} $\bball$ is $(E,\delta_k,K)$-S-good for some $K\in\DN$.
\par\noindent
There exists $L_*(K)\in\DN$ such that if, in addition, $L_0\ge L_*(K)$, then $\bball$ is $(E,m)$-SNS.
\ele
\proof
Derivation of the NS property can be done essentially in the same way as in \cite{DK89} and in numerous
subsequent papers, with minor adaptations. See for example
\lcite{C16b}{proof of Lemma 7} where the singular balls
are also supposed to be pairwise $L_k^\ttau$-distant, $\ttau>1$. 
To show that the  \emph{strong} (stable) non-singularity property also holds true, one can use induction on scales $L_k$.
We have to show that the NS property of the larger cube $\bball_{L_{k+1}}(\Bu)$ is stable with respect to
arbitrary fluctuations of the random amplitudes $\om_y$ with $y\not\in \Pi\bball_{L^\ttau_{k+1}}(\Bu)$.
According to what has just been said in the previous paragraph, it suffices to check the stability of the
properties
\vskip1mm
\noindent
$\mathbf{(i')}$
$\bball_{L_{k+1}}(\Bu)$ is $(E, \eps_k)$-NR,

\vskip1mm
\noindent
$\mathbf{(ii'')}$ \, $\bball_{L_{k+1}}(\Bu)$ contains no collection of cubes $\{\bball_{L_k}(\Bx^i), 1\le i \le S+1\}$,
with pairwise $2L_k^{\ttau}$-distant centers, neither of which is $(E,m)$-NS,

\vskip1mm
\noindent
under the hypotheses \textbf{(i)}--\textbf{(ii)}.

There is nothing to prove for the stability of $\mathbf{(i')}$, as it is asserted by $\mathbf{(i)}$.

On the scale
$L_0$ the non-singularity is derived from non-resonance, with a comfortable gap between an energy $E$
and the spectrum in the cube of radius $L_0$, which provides the base of induction.
Evidently, given any cube
$\bball_{L_{j}}(\Bx) \subset \bball_{L_{k+1}}(\Bu)$ one has
$$
\all j=0, \ldots, k \quad \bball^{\rc}_{L_{k+1}}(\Bu) \subset \bball^{\rc}_{L_{j}}(\Bx) \,.
$$
In other words, stability encoded in the SNS or SNR properties of smaller balls
$\bball_{L_{j}}(\Bx) \subset \bball_{L_{k+1}}(\Bu)$ is stronger than what is required for the stability
w.r.t. fluctuations $\om_y$ outside a much larger cube $\bball_{L^\ttau_{k+1}}(\Bu)$.
We conclude that the claim follows indeed from the the hypotheses \textbf{(i)}--\textbf{(ii)}.
\qedhere

\subsection{Conditions for non-singularity of NI cubes}

\ble
\label{lem:NI.perturb}
Consider a NI cube $\bball = \bball^{(N)}_{L_k}(\Bu) = \ball'\times \ball''$, and the respective reduced Hamiltonians
$H' = H_{\ball'}$ and $H'' = H_{\ball''}$. Assume that $\bcube$ is $(E,2\eps_k)$-SNR and, in addition,
\par\vskip1mm\noindent
$\bullet$ $\all \lam'\in\Sigma(H')$ the cube $\ball''$ is $(E-\lam', \delta_k)$-SNS, and

\par\vskip1mm\noindent
$\bullet$ $\all \lam''\in\Sigma(H'')$ the cube $\ball'$ is $(E-\lam'', \delta_k)$-SNS.
\par\vskip1mm\noindent Then $\bball$ is
$(E,\delta_k)$-SNS.
\ele

The proof of this  deterministic statement is  similar to that of its counterpart from
\cite{CS09a} and subsequent papers on $N$-particle localization with short-range site potentials, except
for the stability aspect. Since the non-singularity of the projection cubes is assumed to be stable (SNS),
so is the resulting non-singularity of the $2$-particle cube $\bball_{L_k}(\Bu)$.

\subsection{Probabilistic analysis}

The following statement is merely an adaptation of Corollary \ref{eq:cor.Wegner.bound},
stated in a form suitable for the fixed-energy scaling analysis.

\ble[Probability of SNR-cubes]
\label{lem:prob.SNR.in.MSA}
For all $k\in\DN$ and $SI$ cubes $\bball_{L_{k}}(\Bu)$ one has
\be
\pr{ \bball_{L_{k}}(\Bu) \text{ is $(E,\eps_{k})$-SNR }} \ge 1 - L_{k}^{ -\frac{(A+c)\ttau}{2}} \,.
\ee
\ele

\vskip3mm

It follows from Definition \ref{def:SNS.SNR.polynom} that any event of the form
$$
 \mcA\big(\bball_L(x), E, m \big) = \myset{ V_\fq(\cdot\,; \om)\big|_{\bball_L(x)} \text{ is $(E,m)$-SNS }  }
$$
is measurable w.r.t. the sigma-algebra $\fF^\fq_{\bball^\ttau_{L}(x)}$.

The next estimate of the probability of occurrence of multiple singular $2$-particle cubes
$\bball_i \equiv \bball_{L_k}(\Bu^{(i)})$, $1\le i \le S_{k+1}$,
is quite similar to its single-particle counterpart, since it treats the case
of distant PI cubes, each located -- by definition -- "closely enough" to the diagonal, so that their full projections
$\Pi \bball_i = \ball_{L_k}(u^{(i)}_1) \cup \ball_{L_k}(u^{(i)}_2)$
are pairwise distant, essentially as in the single-particle case. The main technical difference is that now
we have to control the fluctuations of the locally constant (due to the staircase nature of $\fu$)
random potential on the entire projections $\Pi \bball_i$, while in the $1$-particle systems one has
$\Pi \bball_L(\Bu) \equiv  \bball_{L}(\Bu)$.
In fact, a necessary adaptation was already made in the eigenvalue concentration estimate given in Section \ref{sec:Wegner}.

\ble[Probability of a bad PI-cluster]
\label{lem:factor.prob.S.polynom}
Assume that $A > 2Nd + 3\gamma$ with $\gamma>0$, and $\ttau > 2 + \frac{A}{Nd}$. Set
$\fs = (A - Nd)\ttau - (A+Nd+1)$, $\sigma = \gamma (Nd)^{-1}$ and $\alpha = (1+\sigma)\ttau$. Then
\be
\label{eq:s-alpha.d}
\fs - \alpha Nd > \gamma\ttau >0.
\ee
Further, let $K\in\DN$ satisfy $\frac{2M(1+\sigma)\fs}{\gamma}< K < L_k^{\alpha - \ttau}$, $M\ge 1$. Then for $L_0$ large enough
\be
\label{eq:prob.bound.K.cluster}
\sup_{\Bu\in\DZ^d} \; \pr{\bball_{L_{k+1}}(\Bu) \text{ is not $(E,\delta_k,\ttau,K)$-S-good} } \le p_{k+1}
\le L_{k+1}^{-M\fs}\,.
\ee
\ele

\proof
\eqref{eq:s-alpha.d} follows from the assumptions on $A$ and $\ttau$ by a simple calculation:
$$
\bal
\fs - \alpha Nd
& = (A - Nd)\ttau - (A + Nd + 1) - (1+\sigma)Nd\ttau
\\
&
= \left(\gamma + (\gamma - \sigma Nd) \right)\ttau  - (A + 2Nd)
\\
&
> \gamma\ttau + \big(\gamma\ttau - (A + 2Nd)\big) > \gamma\ttau.
\eal
$$
Further, by Remark \ref{rem:beta}, any collection of events $\mcE_i = \set{\bball_{L_k}(\Bu^i) \text{ is not $(E,m)$-SNS}}$
with pairwise $L_k^{\ttau}$-distant centers $\Bu^i$ is independent, hence
$$
\pr{ \cap_{i=1}^K \mcE_i } = \prod_{i=1}^K \pr{ \mcE_i } \le p_k^K.
$$
By \eqref{eq:s-alpha.d}  we have $\frac{\fs-\alpha Nd}{\alpha} > \frac{\gamma\ttau}{(1+\sigma)\ttau}=\frac{\gamma}{1+\sigma}$.
Thus with $K>2M\fs(1+\sigma)\gamma^{-1}$, we have for the maximal number $\mcS(\om)$ of pairwise $L_k^\ttau$-distant singular
PI cubes $\bball_{L_k^\ttau}(\Bx^i)$ inside $\bball_{L_{k+1}}(\Bu)$:
\be
\pr{ \mcS(\om) \ge K} \lea L_{k+1}^{KNd} p_k^K
\lea L_{k+1}^{ -K\left( \frac{\fs}{\alpha} - Nd\right) }
\le \frac{1}{4} L_{k+1}^{-M\fs}.
\ee
The claim is proved.
\qedhere
\vskip2mm

It is to be stressed that the positive integer $M$ (hence the exponent $M\fs$ of the length scale $L_{k+1}$
in \eqref{eq:prob.bound.K.cluster})
can be made arbitrarily large by taking $L_0$ large enough.

The next statement relies on the $1$-particle localization results for the staircase potentials (cf. \cite{C16f}).
\ble
\label{lem:prob.bad.NI.cube}
If $L_0$ is large enough, then
for any $E\in\DR$ and any NI cube $\bball_{L_k}(\Bu)$ one has
\be
\pr{ \text{$\bball_{L_k}(\Bu)$ is $(E,m)$-S}} \le L_{k+1}^{-\frac{3b}{2}}.
\ee
Consequently, for any $\Bx$
\be
\pr{ \text{$\bball_{L_{k+1}}(\Bx)$ contains a NI $(E,\delta)$-S cube $\bball_{L_k}(\Bu)$}}
\le \frac{1}{4} L_{k+1}^{-b} .
\ee

\ele

The proof is similar to that of \lcite{CS17}{Lemma 3.4}.

\ble[Scaling of probabilities]
\label{lem:ind.prob.polynom}
Consider a Hamiltonian $\BH$ where, as in Lemma \ref{lem:factor.prob.S.polynom}, the site potential
$\fu(r) = r^{-A}$, $A>2Nd+3\gamma$, $\gamma>0$. Let $\fs = (A-Nd)\ttau-(A+Nd+1)$, and assume that the
scale growth exponent (cf. \eqref{eq:con.alpha.tau.S}) has the form $\alpha = (1+\sigma)\ttau$ with $\sigma = \gamma/(Nd)$.
Fix an arbitrarily large $b>0$ and assume that
$$
\sup_{\Bu\in\DZ^{Nd}} \; \pr{\bball_{L_k}(\Bu) \text{ is not $(E,m)$-SNS} } \le p_k \le L_k^{-b} \,.
$$
Furthermore, let the cluster cardinality parameter $K$ in the definition of $(E,\delta,K,\ttau)$-good cubes
satisfy $K > \frac{2M(1+\sigma)}{\gamma}$, where $\DN\ni M > \frac{2b}{\fs}$.
Then
$$
\sup_{\Bu\in\DZ^d} \; \pr{\bball_{L_{k+1}}(\Bu) \text{ is not $(E,m)$-SNS} } \le p_{k+1} \le L_{k+1}^{-b}\,.
$$
\ele

\proof
By Lemma \ref{lem:cond.SNS.polynom}, if $\bball_{L_{k+1}}(u)$ is not $(E,m)$-SNS, then
\begin{enumerate}[\bf(i)]
  \item either it is not $(E,\eps_{k+1})$-SCNR,

  \item or it contains a $(E,m)$-singular, non-interactive  cube $\bball_{L_k}(\By)$,

  \item or it is not $(E,\delta_k,\ttau,K)$-S-good.
\end{enumerate}

\vskip2mm
\noindent
$\bullet$ The probability of the event \textbf{(i)} is assessed with the help of the Wegner-type estimate
from Theorem  \ref{thm:Wegner}, relying
on the disorder in the cubes $\bball_{L^{\ttau}}(\Bu)$ with $\bball_{L}(\Bu)\subseteq \bball_{L_{k+1}}(\Bu)$.
The largest of these cubes, $\bball_{L_{k+1}}(\Bu)$, is surrounded by a belt of width $L_{k+1}^{\ttau}$
where the random amplitudes are not fixed hence can contribute to the Wegner estimate with
$\eps = L_{k+1}^{-A\ttau}$,
hence the same is true for all of these balls:  we have (cf. \eqref{eq:def.btau.kap} and \eqref{eq:cor.Wegner.bound})
\be
\label{eq:prob.SNR.polynom}
\pr{ \bball_{L^{\ttau}}(\Bu) \text{ is not $(E, \eps_R)$-SNR} }
\le L_{k+1}^{-\half A\ttau} \,.
\ee
Since $\ttau>2(b+1)/A$, the RHS of \eqref{eq:prob.SNR.polynom} is bounded by
$\half L_{k+1}^{-b'-1}$ with $b'>b$.
The total number of such
balls is $Y_{k+1} = L_k^{\alpha-1} = L_{k+1}^{1 - \alpha^{-1}}$, with $1 - \alpha^{-1} < 1$,
whence
\begin{align}
\notag
\pr{ \bball_{L_{k+1}^{1+\ttau}}(u) \text{ is not $(E, \eps_R)$-CNR} }
&
\lea L_{k+1}^{-(b'+1) + 1} <  \frac{1}{4} L_{k+1}^{-b } .
\end{align}

\noindent
$\bullet$ Next, by Lemma \ref{lem:prob.bad.NI.cube}, the probability of \textbf{(ii)}
is upper-bounded by $\frac{1}{4} L_{k+1}^{-b}$.

\vskip2mm

\noindent
$\bullet$ To assess the probability of \textbf{(iii)}, recall that by Lemma \ref{lem:factor.prob.S.polynom}
$$
\bal
\prob{ \mcS_{k+1} >  K } &\le \frac{1}{4} L_{k+1}^{-M\fs} \,,
\eal
$$
whence
$$
\pr{  \bball_{L_{k+1}}(u) \text{ is not $(E,m)$-SNS }}
\le \half L_{k+1}^{-b} + \half L_{k+1}^{-b} =  L_{k+1}^{-b} \,.
$$
Collecting the above three estimates, the claim follows.
$\,$
\qedhere

By induction on $k$, we come to the conclusion of the fixed-energy MSA
under a polynomially decaying interaction.

\btm
\label{thm:FEMSA.claim}
Suppose that the ILS estimate
\be
\label{eq:thm.FEMSA.ILS}
\sup_{u\in\DZ^d} \; \pr{  \bball_{L_0}(u) \text{ is not $(E,m)$-SNS } } \le L_0^{-b}
\ee
holds for some $L_0$ large enough, uniformly in $E\in I_* \subset\DR$. Then for all $k\ge 0$
and all $E\in I$
\be
\label{eq:thm.FEMSA.claim}
\sup_{u\in\DZ^d} \;
\pr{  \bball_{L_k}(u) \text{ is not $(E,m)$-SNS } } \le L_k^{-b}.
\ee
\etm

The required ILS estimate is established in Section \ref{sec:ILS}.

This concludes the fixed-energy MSA induction.

\section{Derivation of spectral and dynamical localization}

\subsection{Energy-interval estimates}

\vskip2mm
Proposition \ref{prop:ETV.altern}, proved in \cite{ C14a}, provides
an alternative to an earlier method developed by Elgart \etal \cite{ETV10},
and Proposition \ref{prop:EFC.GK01} is essentially a reformulation
of an argument by Germinet and Klein (cf. \lcite{GK01}{proof of Theorem 3.8}) which substantially
simplified the derivation of strong dynamical localization from the energy-interval MSA bounds,
compared to \cite{DamSto01}.

Introduce the following notation: given a cube $\bball_L(z)$ and $E\in\DR$,
$$
\bal
\BF_{z,L}(E) &:= \big| \bball_L(z) \big| \; \max_{|y-z|} \big| G_{\bball_L(z)}(z,y;E) \big|\,,
\eal
$$
with the convention that $\big| G_{\bball_L(z)}(z,y;E) \big| = +\infty$ if
$E$ is in the spectrum of $H_{\bball_L(z)}$. Further, for a pair of balls
$\bball_L(x), \bball_L(x)$ set
$$
\bal
\BF_{x,y,L}(E) &:= \max\, \big[ \BF_{x,L}(E), \, \BF_{y,L}(E) \big] \,.
\eal
$$

The fixed-energy MSA in an interval $E\in I\subset \DR$ provides probabilistic bounds on
the functional $\BF_{x,L}(E)$ of the operator $H_{\bball_L(x)}(\om)$; as a rule, they are easier
to obtain that those on $\sup_{E\in I} \; \BF_{x,y,L}(E)$ (referred to as energy-interval bounds).
Martinelli and Scoppola \cite{MS85} were apparently the first to notice a relation between the two
kinds of bounds, and used it to prove a.s. absence of a.c. spectrum for Anderson Hamiltonians
obeying suitable fixed-energy bounds on fast decay of their Green functions. Elgart, Tautenhahn and
Veseli\'c \cite{ETV10} improved the Martinelli--Scoppola technique, so that energy-interval bounds
implying spectral and dynamical localization could be derived from the outcome of the fixed-energy MSA.

\bpr[Cf. \lcite{C14a}{Theorem 4}]
\label{prop:ETV.altern}
Let be given a random ensemble of operators $H(\om)$ acting in a finite-dimensional Hilbert space $\mcH$,
$\dim \mcH = D$, two subspaces $\mcH', \mcH''\subset \mcH$ with their respective orthonormal bases
$\set{ \phi_i, 1\le i \le D' }$ and $\set{ \psi_j, 1\le j \le D'' }$, an interval $I\subset\DR$ and real numbers $a, \fp>0$
such that for all $E\in I$ the function
$$
\mcM:\, (E,\om) \mapsto \max_i \; \max_j \; \norm{ \Pi_{\phi_i} G(E,\om) \Pi \psi_j}
$$
(with $\Pi_\phi \equiv \ket{\phi}\bra{\phi}$) satisfies
\be
\pr{ \mcM(E,\om) \ge a} \le \fp.
\ee

Then the following holds true:

\par\vskip1mm\noindent
(A)
For any $b>\fp$ there exists an event $\mcB(b)\subset\Om$ such that $\pr{\mcB(b)} \le b^{-1} \fp$ and
for any $\om\not\in \mcB(b)$ the random set of energies
$$
\cE(a,\om) := \set{E\in I: \, \mcM(E,\om) \ge a}
$$
is covered by $K< 3n' n'' N \le 3N^4$ intervals $J_k = [E^-_k, E^+_k]$ of total length $\sum_k |J_k|\le b$.

\par\vskip1mm\noindent
(B)
The random endpoints $E_k^\pm$ depend upon $H$ in such a way that, for a one-parameter family
$A(t) := H(\om) + t\one$, the endpoints $E_k^\pm(t)$ for the operators $A(t)$ (replacing $H$) have the form
\beal
E_k^\pm(t) &= E_k^\pm(0) + t, \;\; t\in\DR.
\eeal
\epr

For our purposes, it suffices to set $b = \fp^{1/2}$ in assertion (A).

The next statement is an adaptation of \lcite{C16b}{Theorem 6}.

\btm
\label{thm:ETV.2p}
Consider two $2$-particle cubes $\bball' = \bball_L(\Bx)$, $\bball'' = \bball_L(\By)$, and introduce the functions
\beal
\mcM_\Bx(E,\om) &= \max_{\Bz\in \pt^- \bball_L(\Bx)} \abs{ \BG_{\bball_L(\Bx)}(\Bx, \Bz; E)} \,,
\\
\mcM_\By(E,\om) &= \max_{\Bz\in \pt^- \bball_L(\By)} \abs{ \BG_{\bball_L(\By)}(\By, \Bz; E)} \,,
\eeal

Then
\be
\pr{ \exist E\in I:\, \min\big[ \mcM_\Bx(E,\om)\,, \, \mcM_\By(E,\om) \big] \ge a}
\lea  |\bball_L(\Bx)|\, |\bball_L(\By)|\, L^{A{\tiny \btau}} \fp^{1/2}.
\ee
\etm

The proof follows essentially the same path as in  \lcite{C14a}{Proof of Theorem 5},
and is even slightly simpler, for it uses a representation
\beal
\BV(\Bz, \om) \one_{\ball_L(\Bx)} &= \xi(\om) \one_{\ball_L(\Bx)} + \BV'(\Bz,\om)\one_{\ball_L(\Bx)}\,,
\\
\BV(\Bz, \om) \one_{\ball_L(\Bx)} &= c \xi(\om) \one_{\ball_L(\By)}  + \BV''(\Bz,\om)\one_{\ball_L(\By)}\,,
\eeal
where  $\BV'(\Bz,\om)$ and $\BV''(\Bz,\om)$ are measurable with respect to some sub-sigma-algebra $\fB\subset \fF$,
while $\xi(\cdot)$ is independent  of $\fB$ and has the continuity modulus $\fs_\xi$.
Such a representation was obtained in Section \ref{sec:EV.compare.staircase}. In \cite{C16b} one had to assess
first the (random) continuity modulus of the conditional sample mean of the random potential in a finite cube.

\subsection{Decay of eigenfunction correlators}

Given an interval $I\subset \DR$,
denote by $\cscB_1(I)$ the set of  bounded Borel functions $\phi:\, \DR\to\DC$ with
$\supp\, \phi\subset I$ and $\| \phi\|_\infty \le 1$.

\bpr[Cf. \lcite{C16a}{Theorem 3}, \cite{GK01}]
\label{prop:EFC.GK01}
Assume that the following bound holds for some $\eps>0$, $\rh_L>0$,
$L\in\DN$ and  a pair of balls $\bball_{L}(x), \bball_{L}(y)$ with $|x-y|\ge 2L+1$:
\be
\label{prop:cond.EFC.GK01}
\;  \pr{\sup_{E\in I} \; \min_{\Bz\in\set{\Bx, \By} }\mcM_{\Bz}(E,\om) > \eps} \le \rh_L.
\ee
Then for any cube $\bball \supset \big(\bball_{L+1}(x) \cup \bball_{L+1}(y) \big)$
\be
\label{prop:EFC.GK01.claim}
\esm{ \sup_{\phi\in\cscB_1(I)} \big| \lr{ \one_x \,|\, \phi\big(H_\bball\big) \, \one_y \,} }
\le 4\eps + \rh_L \,.
\ee
\epr

\proof[Proof of assertion (B), Theorem \ref{thm:main.polynom.low.E}]
The validity of the condition \eqref{prop:cond.EFC.GK01}
with $\rh_L = L^{-\frac{1}{4}\fb + C(A)}$ follows from Theorem \ref{thm:ETV.2p}.
Since $\fb>0$ can be made arbitrarily large by taking $L_0$ and the auxiliary parameter
$\ttau$ large enough, the assertion (B) of Theorem \ref{thm:main.polynom.low.E} follows.
\qedhere



\begin{thebibliography}{10}
\providecommand{\url}[1]{{#1}}
\providecommand{\urlprefix}{URL }
\expandafter\ifx\csname urlstyle\endcsname\relax
  \providecommand{\doi}[1]{DOI~\discretionary{}{}{}#1}\else
  \providecommand{\doi}{DOI~\discretionary{}{}{}\begingroup
  \urlstyle{rm}\Url}\fi

\bibitem{AW09}
Aizenman, M., Warzel, S.: Localization bounds for multiparticle systems.
\newblock Commun. Math. Phys. \textbf{290}, 903--934 (2009)

\bibitem{BK05}
Bourgain, J., Kenig, W.: On localization in the continuous
  {A}nderson--{B}ernoulli model in higher dimension.
\newblock Invent. Math. \textbf{161}, 389--426 (2005)

\bibitem{C16d}
Chulaevs{k}y, V.: Renormalization group limit of {A}nderson models.
\newblock submitted

\bibitem{C14a}
Chul{a}e{v}{s}ky, V.: From fixed-energy localization analysis to dynamical
  localization: An elementary path.
\newblock J. Stat. Phys. \textbf{154}, 1391--1429 (2014)

\bibitem{C16a}
Chulaev{s}ky, V.: Exponential scaling limit for single-particle {A}nderson
  models via adaptive feedback scaling.
\newblock J. Stat. Phys. \textbf{162}, 603--614 (2016)


\bibitem{C16b}
Chulaevs{k}y, V.: Efficient localization bounds in a continuous {N}-particle
  {A}nderson model with long-range interaction.
\newblock Lett. Math. Phys. \textbf{106}(4), 509--533 (2016)


\bibitem{C16e}
C{h}ulaevsky, V.: Universality of smoothness of {D}ensity of {S}tates in
  arbitrary higher-dimensional disorder under non-local interactions {I}.
  {F}rom {V}i\'{e}te--{E}uler identity to {A}nderson localization.
\newblock \texttt{arXiv:math-ph/1604.08534} (2016)


\bibitem{C16f}
Ch{u}laevsky, V.: Density of states under non-local interactions {II}.
  {S}implified polynomially screened interactions.
\newblock Preprint, \texttt{arXiv:math-ph/1606.05618} (2016)

\bibitem{CBS11}
Chulaevsky, V., Boutet~de Monvel, A., Suhov, Y.: Dynamical localization for a
  multi-particle model with an alloy-type external random potential.
\newblock Nonlinearity \textbf{24}(5), 1451--1472 (2011)

\bibitem{CS09a}
Chulaevsky, V., S{u}hov, Y.: Eigenfunctions in a two-particle {A}nderson tight
  binding model.
\newblock Commun. Math. Phys. \textbf{289}, 701--723 (2009)

\bibitem{CS09b}
Chulaevsky, V., S{u}hov, Y.: Multi-particle {A}nderson localisation: Induction
  on the number of particles.
\newblock Math. Phys. Anal. Geom. \textbf{12}, 117--139 (2009)

\bibitem{CS17}
Chulaevsky, V., Suh{o}v, Y.: Efficient {A}nderson localization bounds for large
  multi-partilce systems.
\newblock J. Spec. Theory \textbf{7}, 269--320 (2017)

\bibitem{DamSto01}
Damanik, D., Stollmann, P.: Multi-scale analysis implies strong dynamical
  localization.
\newblock Geom. Funct. Anal. \textbf{11}(1), 11--29 (2001)

\bibitem{DK89}
von Dreifus, H., Klein, A.: A new proof of localization in the {A}nderson tight
  binding model.
\newblock Commun. Math. Phys. \textbf{124}, 285--299 (1989)

\bibitem{ETV10}
Elgart, A., Tautenhahn, M., Veseli\'c, I.: Localization via fractional moments
  for models on \textbf{Z} with single-site potentials of finite support.
\newblock J. Phys. {A} \textbf{43}(8), 474,021 (2010)

\bibitem{FW15}
Fauser, M., Warzel, S.: Multiparticle localization for disordered systems on
  continuous space via the fractional moment method.
\newblock Rev. Math. Phys. \textbf{27}(4), {1550,010} (2015)

\bibitem{FMSS85}
Fr\"{o}hlich, J., Martinelli, F., Scoppola, E., Spencer, T.: Constructive proof
  of localization in the {A}nderson tight binding model.
\newblock Commun. Math. Phys. \textbf{101}, 21--46 (1985)

\bibitem{FS83}
Fr\"{o}hlich, J., Spencer, T.: Absence of diffusion in the {A}nderson tight
  binding model for large disorder or low energy.
\newblock Commun. Math. Phys. \textbf{88}, 151--184 (1983)

\bibitem{GK01}
Germinet, F., Klein, A.: Bootstrap multi-scale analysis and localization in
  random media.
\newblock Commun. Math. Phys. \textbf{222}, 415--448 (2001)

\bibitem{KirStoStolz98a}
Kirsch, W., Stollmann, P., Stolz, G.: Anderson localization for random
  {S}chr\"{o}dinger operators with long range interactions.
\newblock Commun. Math. Phys. \textbf{195}, 495--507 (1998)

\bibitem{KN13}
Klein, A., Nguyen, S.T.: Bootstrap multiscale analysis for the multi-particle
  {A}nderson model.
\newblock J. Stat. Phys. \textbf{151}(5), 938--973 (2013)

\bibitem{KN16}
Klein, A., N{g}uyen, S.T.: Bootstrap multiscale analysis for the multi-particle
  continuous {A}nderson {H}amiltonians.
\newblock J. Spec. Theory \textbf{5}(2), 399--444 (2016)

\bibitem{MS85}
Martinelli, F., Scoppola, E.: Remark on the absence of absolutely continuous
  spectrum for $d$-dimensional {S}chr\"{o}dinger operators with random
  potential for large disorder or low energy.
\newblock Commun. Math. Phys. \textbf{97}, 465--471 (1985)

\bibitem{Wint1934}
Wintner, A.: On analytic convolutions of {B}ernoulli distributions.
\newblock Amer. J. Math. \textbf{56}, 659--663 (1934)

\end{thebibliography}



\end{document}